\documentstyle[aps,prl,epsf]{revtex}
\begin{document}
\twocolumn[\hsize\textwidth\columnwidth\hsize\csname@twocolumnfalse\endcsname

\title{Critical Current Enhancement due to an Electric Field in a 
Granular $d$-Wave Superconductor}
\author{Daniel Dom\'{\i}nguez, Cristina Wiecko}
\address{Centro At\'{o}mico Bariloche, 8400 S. C. de Bariloche, 
Rio Negro, Argentina.}
\author{Jorge V. Jos\'{e}}
\address{Physics Department and Center for the Interdisciplinary
Research on Complex Systems \\ Northeastern University, Boston, MA,
02115, USA.}
\date{Phys. Rev. Lett. Nov 22, 1999}
\draft
\maketitle
\begin{abstract}

We study the effects of  an electric-field in the transport
properties of bulk granular superconductors with different
kinds of disorder. We find that for
a $d$-wave granular superconductor with random $\pi$-junctions
the critical current always increases after applying a strong electric
field, regardless of the polarity of the field. This result plus a change in the
voltage as a function of the electric field are in good agreement with
experimental results in ceramic high $T_c$ superconductors.

\end{abstract}

] 

\narrowtext

Recent experiments in ceramic high-$T_c$ superconductors (HTCS) 
have found an {\it enhancement} in the critical current 
when applying an electric field $E$ through an insulating layer 
~\cite{sko92,sok93,os94,more}. Previous studies of the
electric field effects (EFE) in superconducting films  have 
attributed the changes in the critical current to variations in the
charge density or to a redistribution of carriers,
 which appear at the surface layer with
depths of the order of the electrostatic screening length 
$d_E$ (in HTCS, $d_E\approx 5\, \AA$) \cite{reviews}.
Several experiments in ultrathin YBa$_2$Cu$_3$O$_{7-x}$ films 
($5-10\,$nm thick) have found 
that $E$  can affect $T_c$ and the $IV$ characteristics
\cite{films}, in good agreement with this picture \cite{reviews}.
In this case, there is either an enhancement or a 
depletion of the critical current
depending on the polarity of $E$ \cite{films}.
The surprising observation in \cite{sko92,sok93,os94,more} 
of a strong EFE in {\it bulk} ceramic HTCS ($1.5$ mm thick), however,
can not be explained  by a surface effect. 
Moreover, for high enough electric fields, the critical current  
{\it always increases}  regardless of the polarity of the field \cite{os94}.

Rakhmanov and Rozhkov \cite{rak}  have shown that an electric field 
can induce a change in the critical currents of the Josephson junctions
present in granular samples.
However, in their model the critical current either increases or decreases
depending on the sign of $E$. Recently, Sergeenkov and Jos\'{e} \cite{serg}
have proposed  that an electric field applied to a 
granular superconductor can produce a magneto-electric like effect,
which could be indirectly related to the behavior of 
the critical current in  \cite{sko92,sok93,os94,more}, 
but no comparison  with the experimental results was given.

Granular superconductors are usually described as 
a random network of superconducting grains  coupled by Josephson
weak links \cite{ebner,choi}. In the HTCS ceramics
several experimental groups have found a paramagnetic Meissner effect (PME) 
at low magnetic fields \cite{pme}. 
Sigrist and Rice \cite{sigrist} proposed that this 
effect could be a consequence of the
intrinsic unconventional pairing symmetry of the HTCS of $d_{x^2-y^2}$ type
\cite{wollman}. Depending
on the relative orientation of the superconducting grains, it is possible
to have  weak links with negative Josephson coupling 
($\pi$-junctions) \cite{sigrist,wollman} 
which, according to \cite{pme,sigrist}, give rise to the PME \cite{niobio}. 
In this paper we will show that the presence of $\pi$-junctions 
in ceramic samples also explains the unusual electric field effects 
observed in \cite{sko92,sok93,os94,more}.

We consider a 3-D cubic network of superconducting grains \cite{ebner,choi} 
at ${\bf n}=(n_x,n_y,n_z)$  and with unit vectors
${\hat\mu}={\hat x},{\hat y},{\hat z}$. The current $I_{\hat\mu}({\bf n})$  
between two grains ${\bf n}$, and ${\bf n}+\hat\mu$  is given
by the sum of the Josephson supercurrent plus a dissipative  Ohmic current:
\begin{equation}
I_{\mu}({\bf n})= I^0_{{\bf n},\mu}\sin\theta_\mu({\bf n})
+\frac{\Phi_{0}}{c2\pi{R}}\frac{d \theta_\mu({\bf n})}{dt}
\;. 
\end{equation}
Here $\theta_\mu({\bf n})=\theta({\bf n}+\hat\mu)-\theta({\bf n})-
A_{\mu}({\bf n},t)$ is the gauge invariant phase difference,
with $\theta({\bf n})$ the superconducting phase in each grain,
and $A_{\mu}({\bf n},t)=
\frac{2\pi}{\Phi_0}\int_{\bf n}^{{\bf n}+\hat\mu}{\bf A}\cdot d{\bf l}$ 
($\Phi_0=h/2e$).
 The critical current of each junction is  $I^0_{{\bf n},\mu}$ and 
${R}$ is the normal state tunneling resistance between grains. 
Together with the conditions of current conservation,
$ \sum_{\mu} [I_{\mu}({\bf n})-I_{\mu}({\bf n}-\hat \mu)]=0$,
 this determines the dynamical equations for the Josephson network. 
We consider periodic boundary conditions (PBC) in a network 
with $N\times N\times N$ grains.

When an electric field ${\bf E}$ is applied 
in the ${\hat z}$ direction, the $z$ component of the vector potential
is given by:
\begin{equation}
A_z({\bf n},t)=A_z({\bf n},0) - \frac{2 \pi cd}{\Phi_0} E  t,
\end{equation}
with $d$ the intergrain distance or junction thickness. This results
in a high-frequency alternating supercurrent in the $z$-direction
due to the ac Josephson effect \cite{barone}.
In addition, we consider that the sample is driven 
by an external current density $I_{ext}$ along the ${\hat y}$ direction.
Therefore, the vector potential term is
$A_{\mu}({\bf n},t)=-\delta_{\mu,z}\omega_Et-\delta_{\mu,y}\alpha_y(t)$,
where the electric field frequency is $\omega_E=2\pi c Ed/\Phi_0$,
and we consider that no external magnetic field is applied.
(The experiments had zero magnetic field \cite{sko92,sok93,os94,more}). 
The external current density
with PBC determines the dynamics of $\alpha_y(t)$ as
\begin{equation}
I_{ext}=
\frac{1}{N^3}\sum_{\bf n}
 I^0_{{\bf n},y}\sin\theta_y({\bf n})
+\frac{\Phi_{0}}{c2\pi{R}}\frac{d\alpha_y}{dt}\;.
\end{equation}
The average voltage per junction induced by the driving current is then obtained
by $V=\frac{\Phi_0}{2\pi c}\langle\frac{d\alpha_y}{dt}\rangle$.

Furthermore, we consider that the applied electric field is screened inside the 
grains and acting only in the insulating intergrain region of the junctions, of typical 
thickness $d\sim 10-20 \AA$. We also neglect the effects
of intergrain capacitance $C_i$ and intragrain capacitance $C_g$.
In general, the effect of the capacitances is to screen the applied
field $E_{ext}$ inside the sample within an overall screening length
$\lambda_E\sim(C_i/C_g)^{1/2}$. Since $C_g\ll C_i$, $\lambda_E$ is very
large and we can consider that the internal field 
acting in the intergrain junctions is $E_{in}=E\approx\alpha_p E_{ext}$
with the polarizability $\alpha_p \lesssim 1$. 
 We also neglect the possible $E$ dependence of the 
critical currents of the junctions. 
As shown in \cite{rak} this assumption may result in an increase or decrease
of $I^0_{{\bf n},\mu}$ depending on the sign of $E_{in}$. 
We also neglect screening current effects (i.e. finite self-inductances). 
The self-induced magnetic fields are very important for the description
of the critical state  \cite{induct}  
{\it at large magnetic fields},
and for the PME at low magnetic fields \cite{pme_nos}. 
The experiments of \cite{sko92,sok93,os94,more} 
were done at {\it zero} magnetic field and finite electric fields. Thus  
the most important approximation we make is to neglect
the capacitances (i.e. the screening of ${\bf E}$).
After making all these physical approximations, we  will now
focus on the {\it collective} aspects of  the ac Josephson effect  
induced by the electric field.  In this model, the 
electric field scale is given by 
$E_0=RI_0(T=0)/d=\pi\Delta_s(0)/2ed$, with $\Delta_s(0)$ 
the superconducting energy gap; for the YBaCu0 ceramics we have 
$\Delta_s(0)\approx 20 meV$, which gives $E_0\sim 30 MV/m$, i.e. in the
same range as the fields used in  \cite{sko92,sok93,os94,more}.

We assume that the effect of disorder in a granular superconductor
at zero magnetic field mainly modifies the magnitude of the critical currents.
In $s$-wave superconductors the sign of the Josephson coupling is always 
positive. In $d$-wave ceramics
the sign of $I^0_{{\bf n},\mu}$ is expected to vary randomly depending on
the relative spatial orientation of the grains.
Therefore, we consider two models of disorder:  
(i) Granular $s$-wave superconductor (GsS):
We consider $I^0_{{\bf n},\mu}$ 
a random variable with a uniform distribution in the interval 
$[I_0(1-\Delta_c),I_0(1+\Delta_c)]$  
with $\langle I^0_{{\bf n},\mu}\rangle=I_0$ and $\Delta_c<1$.
(ii) Granular $d$-wave superconductors (GdS): We consider that there
is a random concentration $c$ of $\pi$-junctions with
$I^0_{{\bf n},\mu}=-I_0$,  
and $I^0_{{\bf n},\mu}=I_0$ with concentration $1-c$
\cite{pme_nos,pme_abs,kawamura}. The latter model was used to explain
the history-dependent paramagnetic effect \cite{pme_nos}, the anomalous
microwave absorption \cite{pme_abs} and the non-linear susceptibility
and glassy behavior \cite{kawamura} observed experimentally  
in ceramic HTCS \cite{pme}.
  
We have performed numerical simulations of 
systems of sizes $N=8,16$ with a numerical integration of
the dynamical equations with time step  $\Delta t= 0.1 \tau_J$, 
with $\tau_J=\Phi_0/2\pi c RI_0$, and we took $5 \times 10^4$ 
integration steps.  We calculated the current-voltage characteristics 
(IV) for different amounts of disorder and electric fields, averaging
over 20 realizations of disorder in each case.

In the absence of disorder, the IV curves are unaffected by the
electric field. In a perfect cubic network, a finite electric field
induces an ac supercurrent $I_0\sin(\omega_Et)$
along the $z$ axis only, and therefore
the IV curves in the $xy$ plane are not affected by the value of $E$.
When the disorder is small, the amplitude of
the ac supercurrents in the $z$ direction is random and, due to
current conservation in each node of the network, this induces
small ac-currents in the $xy$ planes with random amplitudes
 $\sim \Delta_c I_0\sin(\omega_Et)$.
Adding a small ac current to a Josephson junction {\it reduces} 
its effective dc critical current \cite{barone}. 
In fact, this is the effect we observe in the GsS model.
In Fig.~1(a) we show the IV curve for $\Delta_c=0.6$. There we see that after
applying an electric field $E$ the whole IV curve shifts to lower 
values of the current, the apparent critical current decreases, and therefore
the voltage change $\Delta V=V(E)-V(0)$ is positive for a given current $I>I_c$.
However, this is in the opposite direction of the experimental
results  of  \cite{sko92,sok93,os94,more}, 
where an {\it increase} in the critical current  was observed.

\begin{figure}
\centerline{\epsfxsize=8.0cm \epsfbox{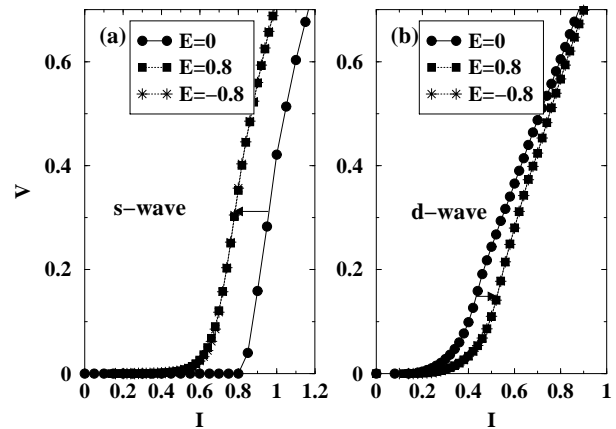}}
\caption{Current-voltage characteristics before and after applying an electric
field $E$. (a) For a granular $s$-wave superconductor (GsS), 
with $\Delta_c=0.6$.
(b) For a granular $d$-wave superconductor (GdS)  with concentration
$c=0.5$ of $\pi$-junctions. 
Voltages are normalized by $NRI_0$ and currents by $N^2I_0$ and lattice
sizes are $16\times16\times16$. 
}
\end{figure}

Let us then consider the case of a granular $d$-wave superconductor.
The presence of randomly distributed negative and positive critical currents
changes the previous simplistic picture, since now 
frustration is introduced in each loop containing an odd number of
$\pi$-junctions \cite{sigrist,pme_nos,kawamura}.
In Fig.~1(b) we show the IV curve for the GdS model 
with a concentration  $c=0.5$
of $\pi$-junctions. We note that after applying the electric field
the IV curve is shifted towards higher currents, i.e. the 
``apparent'' critical current {\it increases}. 
If we change  $E \rightarrow -E$ the IV
curves overlap showing that the effect is independent of the polarity of
the electric field. A similar effect was seen in 
the experiments of \cite{os94}: the IV curves shift
upwards for large $E$ and almost overlap when changing 
the polarity of the field \cite{comment}.
 This result is  surprising since in the usual electric field effect observed 
in films  a change in the polarity of the field drastically
changes the sign of the shift of the IV curves \cite{films}.
 We also see in Fig.1(b) that the EFE is stronger 
 near the critical current, and for large currents 
 the change in voltage $\Delta V$ is smaller, which was 
also seen in experiment \cite{sko92,sok93,os94,more}.
If the amount of disorder in the GsS is such that $\Delta_c > 1$, 
then a fraction $(\Delta_c-1)/2\Delta_c$ of the critical currents will 
be negative. This case is unrealistic for $s$-wave superconductors, but
corresponds to a $d$-wave granular superconductor with a
small concentration of $\pi$-junctions. In fact, we also  
find an increase of the critical current for $\Delta_c>1$.

\begin{figure}
\centerline{\epsfxsize=8.0cm \epsfbox{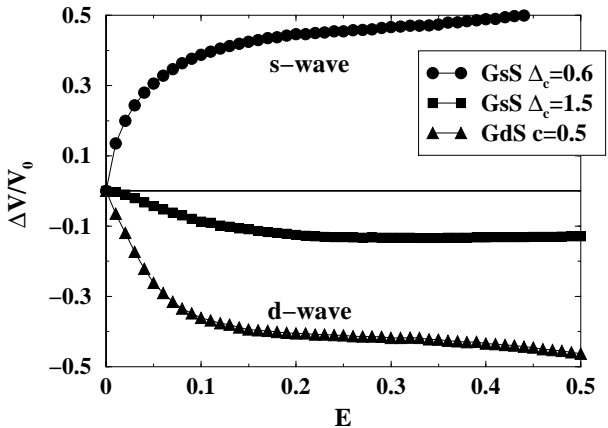}}
\caption{Changes in voltages as a function of electric field for
different kinds of disorder. $\Delta V/V_0= (V(E)-V(0))/V(0)$. The results
for the GsS model are for $I=0.9I_0$ and
for the GdS model, $I=0.5I_0$.
Electric fields are normalized by $E_0$.}
\end{figure}

In Fig.~2 we study the relative change in voltage $\Delta
V/V(0)$ for a fixed bias current (above the critical current)
as a function of $E$. We see that for the GsS model
$\Delta V$ is positive, and only when negative critical currents
are allowed ($\Delta_c>1$), 
there is a decrease in the voltage, i.e. $\Delta V < 0$. The effect
is stronger in the GdS model with  $c=0.5$ concentration of 
$\pi$-junctions. An effect of the same order, i.e. near a $50 \%$ decrease
in voltage, and in the same scale of electric fields was
observed in experiment.
In Fig.~3 we show the dependence of the EFE with different amounts of disorder.
For a given bias current we calculate
$\Delta V/V_0$ after applying a large electric field of $E=0.8E_0$. 
We see in Fig.~3(a) the voltage change in the GsS model as a function
of $\Delta_c$. It is clear that only when $\Delta_c > 1$ the $\Delta V$
can be negative. In Fig.~3(b) we study the same situation for the
GdS model as function of the concentration of the $\pi$-junctions.
We obtain that a small amount of $\pi$-junctions is enough to
show a strong decrease in voltage at large electric fields. 
We see that there is
a very rapid change in $\Delta V/V_0$ close to $c=0.1$. 

\begin{figure}
\centerline{\epsfxsize=8.0cm \epsfbox{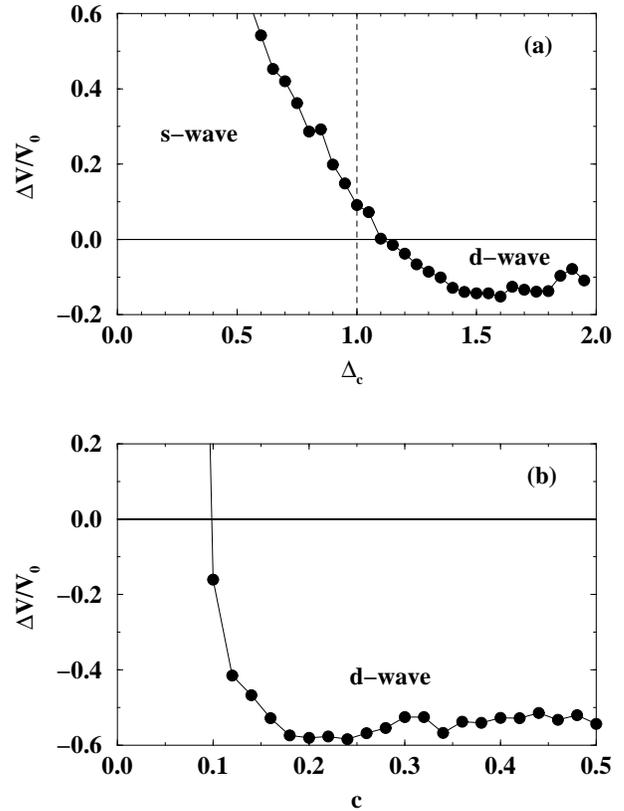}}
\caption{Effects of disorder in the voltage change after applying
an electric field $E$.  (a) GsS model, $\Delta V/V_0$ vs.
$\Delta_c$, $I=0.9I_0$.  (b) GdS model,
$\Delta V/V_0$ vs. $c$, $I=0.5I_0$. 
All the results were obtained with a field 
$E=0.8E_0$.}
\end{figure}

It is interesting to note that  there is another type of disorder that 
can give a negative $\Delta V/V_0$. In the presence of  external magnetic fields, 
the vector potential has a component $A^H_\mu({\bf n})$. For strong 
magnetic fields,  $A^H_\mu({\bf n})$ can be taken as a random variable 
in the interval $[0,2\pi\Delta_H]$, with the case $\Delta_H=1$ corresponding to the 
gauge glass model \cite{gauge}. We have found that only when $\Delta_H>0.5$ 
we can obtain a decrease in the relative voltage, $\Delta V/V_0 < 0$,
similar to the results obtained with the GdS model. The case $\Delta_H>0.5$
is when the effective couplings can take negative values. 
This case, even when interesting in itself,
is unrealistic for this problem  since it corresponds to ceramics at 
very large magnetic fields, while the experiments had zero magnetic field.

There are other interesting aspects of the experimental results. It 
was found that for low electric fields the critical current 
decreases and only for large fields it increases.  
A typical measurement shows that, when applying 
a current near the critical  current, $\Delta V/V_0$ increases  
as a function of $E$ and, after  reaching a maximum, it falls and then
becomes negative for large enough fields.  Our random $d$-wave model for a 
granular superconductor also reproduces this effect in good agreement 
with experiment as we show in Fig.~4. 
We note that the value of $\Delta V/V_0$ at the
maximum decreases when increasing the bias current, and only for 
large enough currents, the voltage change $\Delta V/V_0$ is always negative 
and decreasing with $E$ as previously shown in Fig.~2. The same 
dependence with the
bias current has been observed in experiment (see for example
Figs.3 and Fig.4 of \cite{sko92} and Fig.3 of \cite{sok93}).
These results can be understood by looking at the shape of the
 IV curves at low electric fields. As we show in the inset 
 of Fig.4, for low $E$  there is a crossing of the IV curves, 
 which explains the behavior observed in $\Delta V/V_0$.

\begin{figure}
\centerline{\epsfxsize=8.0cm \epsfbox{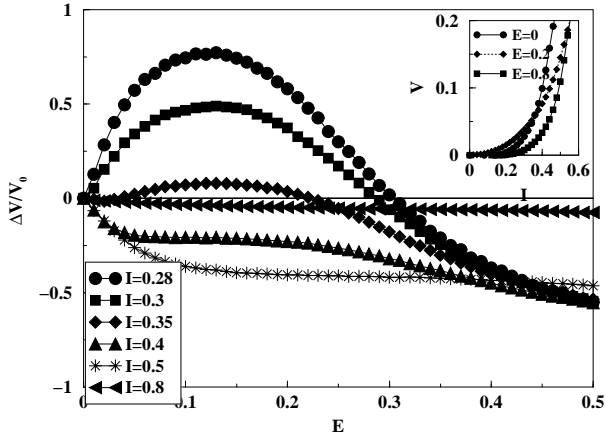}}
\caption{
Change in voltage as a function of $E$ for different driving
currents for the GdS model with $c=0.5$. 
Inset: Crossing of IV curves for different electric
fields.}
\end{figure}
  
The experiments on the electric field effects in 
ceramic HTSC were surprising for two reasons:
(i) a large electric field effect in a bulk ceramic sample was unexpected
\cite{sko92}, (ii) an increase in the apparent critical current as a function 
of $E$ was found to be independent of the polarity of $E$ \cite{os94}. 
Here we have qualitatively explained all  these
experimental features with a simplified model of 
a granular superconductor, which has two basic  ingredients: 
the ac Josephson effect induced by the electric field 
and the collective frustration effects due to $\pi$-junctions.  
There are many interesting open questions such as:
the coexistence of this effect with the paramagnetic Meissner effect
\cite{pme,sigrist,pme_nos},  the time dependent glassy dynamics 
as a function of electric and magnetic fields \cite{os94,more},
the effects in a chiral glass state \cite{kawamura},
and the effect of finite inductances and capacitances in a more 
realistic model. 
We expect  that the results presented here will motivate 
further experimental and theoretical studies of this problem.

This work has been supported by a cooperation grant 
CONICET Res. 697/96 and NSF INT-9602924. 
DD and CW  acknowledge CONICET and CNEA (Argentina) 
for local financial support. The work
by DD was also partially funded by Fundaci\'{o}n Antorchas and ANPCyT 
and the work by JVJ by NSF DMR-9521845.


\begin{thebibliography}{99}
\bibitem{comment} On the other hand, 
in the experiments reported in \cite{sko92} 
the change in voltage $\Delta V$  
is different when changing $E\rightarrow -E$
(but always $\Delta V <0$ for large $E$). This asymmetry could be easily
explained by considering the
dependence with $E$ of the Josephson critical currents $I_0(E)$
\cite{rak}.
\end{thebibliography}
\end{document}